# Positional Dependence of Pulse Shape Discrimination (PSD) in a Monolithic CLLB Crystal

Richard S. Woolf, *Member, IEEE*, Bernard F. Phlips, Anthony L. Hutcheson, Andrew D. Maris, and Eric A. Wulf

*Abstract*–We report on the results of the positional-dependent pulse shape discrimination (PSD) parameter observed within a monolithic CLLB scintillation crystal. CLLB, a relatively novel inorganic scintillation crystal, is capable of PSD between gamma rays, neutrons, and alpha particles. In this work, we observed distinguishable differences in the pulse shapes for gamma-ray-induced events. The CLLB crystals used for this experiment are 5 cm (diameter) by 10 cm (length). By using monoenergetic 2.614 MeV photons from a set of thoriated welding rods and performing collimated scans along the length of the crystal, we found that the centroid of the PSD distribution shifted as a function of position. With positional-dependent PSD, one can obtain more accurate knowledge of the interaction location within a monolithic scintillation crystal. These results could lead to improved angular resolution in imaging systems employing scintillation crystals that exhibit this behavior. Lastly, an understanding of the dependence of the PSD as a function of position could give manufacturers a better understanding of the crystal properties and provide insight to the distribution of internal contaminants and dopants within the crystal.

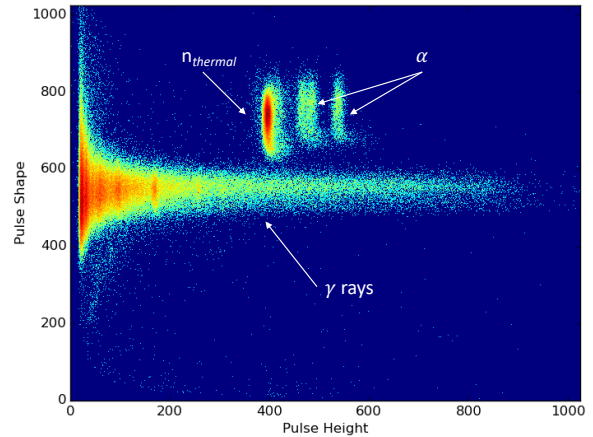

Fig. 1. Pulse shape vs. pulse height for a CLLB detector irradiated by a neutron/gamma-ray source. Identified in the plot are the gamma rays, thermal neutrons and the internal alpha particles.

## I. Introduction

THIS work concentrates on the positional dependence of the pulse shape discrimination (PSD) parameter observed in a novel inorganic scintillation crystal, known chemically as $Cs_2LiLaBr_6$:Ce (CLLB)[1][2]. CLLB is part of the elpasolite family of inorganic scintillation crystals, all of which have the same molecular crystalline structure ($A_2BLnX_6$). Elpasolites are characterized by their high light output and good proportionality, which leads to a scintillation crystal with excellent spectroscopic abilities. With CLLB, <3% at 662 keV (FWHM) has been observed. Additionally, CLLB crystals can detect and discriminate thermal neutrons from gamma rays using the pulse shape discrimination (PSD) technique using the $^6Li(n,\alpha)^3H$ reaction. Moreover, PSD methods can be used to separate gamma rays and thermal neutrons from the internal alpha-particle emission observed in these crystals (Fig. 1).

We have previously reported on the CLLB crystal in terms of the thermal neutron discrimination efficiency in the presence of internal alpha-particle background [3], and investigated the source of internal background contaminants, for CLLB and other inorganic scintillators, by measuring the coincident internal alpha-particle and the associated gamma-ray emission measured by an external detector [4].

## II. Motivation

The main motivation for performing this experiment can be summarized by examining the 2D scatter plots of the pulse shape parameter vs the pulse height (uncalibrated energy) shown in Fig. 2. We obtained these data by irradiating two separate CLLB detectors with a $^{252}$Cf spontaneous fission neutron/gamma-ray source, resulting in the expected horizontal gamma-ray band, the thermal neutron island, and the internal alpha-particle islands. However, a comparison of the thermal neutron (and alpha-particle) islands from these two detectors revealed an energy- and pulse-shape-dependent skewing, present in one detector (Fig. 2, *left*) and not present in the other (Fig. 2, *right*).

Manuscript received December 20, 2020. This work was supported by the Office of Naval Research 6.1.

R. S. Woolf is with the U. S. Naval Research Laboratory, Washington, DC 20375 USA (telephone: 202-404-2886, e-mail: richard.woolf@nrl.navy.mil).
B. F. Phlips is with the U. S. Naval Research Laboratory, Washington, DC 20375 USA (telephone: 202-767-3572, e-mail: phlips@nrl.navy.mil).
A. L. Hutcheson is with the U. S. Naval Research Laboratory, Washington, DC 20375 USA (telephone: 202-404-1464, e-mail: anthony.hutcheson@nrl.navy.mil).
A. D. Maris was with the Naval Research Enterprise Internship Program (NREIP), 1818 N St. NW, Suite 600, Washington, DC 20036 USA, (telephone: 202-404-2886, e-mail: andrew.maris.ctr@nrl.navy.mil).
E. A. Wulf is with the U. S. Naval Research Laboratory, Washington, DC 20375 USA (telephone: 202-404-1475, e-mail: wulf@nrl.navy.mil).

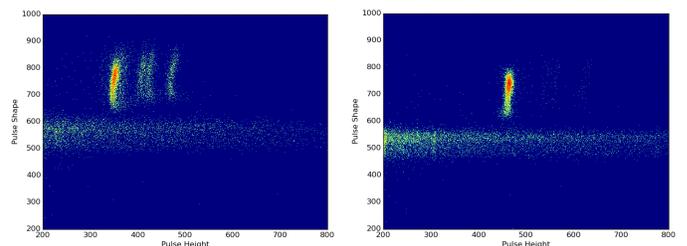

Fig. 2. PSD vs. pulse height for two separate CLLB detectors demonstrating skewing (*left*) and no skewing (*right*) in the neutron/alphas.

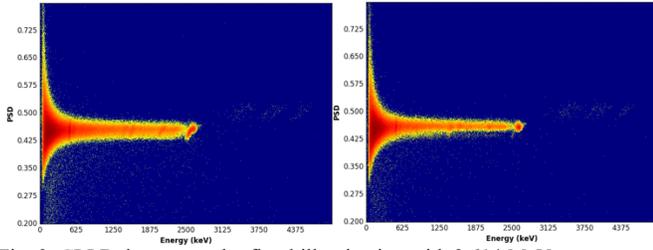

Fig. 3. CLLB detector under flood illumination with 2.614 MeV gamma rays, uncollimated (*left*) and the bottom half shielded by a lead brick (*right*).

We discussed these features with the crystal manufacturer (St. Gobain Crystals) and it was hypothesized that the observed skewing could be the result of an uneven distribution of impurities along the length of the crystal that occurs during the growth process [5].

Based on the results obtained with the $^{252}$Cf source, we investigated the behavior in response to gamma rays using a bright source of 2.614 MeV photons from $^{228}$Th. Fig. 3, *left* shows the skewing effect visible in the gamma-ray line from the CLLB detector under flood illumination. Because of the non-uniform distribution of impurities, which causes the skewing effect observed in thermal neutron/alpha-particle islands, we investigated whether or not "collimating" the source would have an effect on the gamma-ray events. Fig. 3, *right* shows the effect of a crude collimation (i.e., shielding half the detector with a 5-cm-thick lead brick). The effect of this crude collimation is clear in the 2.614 MeV gamma-ray distribution. The events remaining in Fig. 3, *right* are mostly associated with events interacting in the top half of the crystal, while the events shown skewed below the circular island in Figure 3, *left* are associated with events interacting in the bottom half of the crystal. These results led us to perform a systematic scan of CLLB detectors using a narrowly collimated source of 2.614 MeV photons.

### III. EXPERIMENTAL SET UP

In this section we describe the experimental set up used to understand the observed effects of skewing, and ultimately the positional-dependent PSD in the CLLB detectors. We have an array of 100 5 cm (diameter) by 10 cm (length) CLLB detectors in house. These detectors were manufactured by St. Gobain crystals, and to date, are the largest volume CLLB crystals commercially available. The array of CLLB detectors demonstrate an energy resolution (FWHM) ranging from 3.3% to 4.4% at 662 keV.

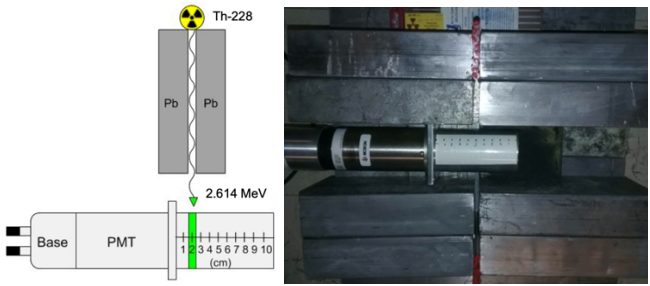

Fig. 4. Cartoon (*left*) and photograph (*right*) depicting the experimental set up used.

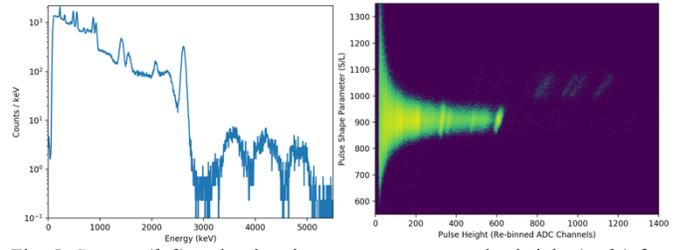

Fig. 5. Spectra (*left*) and pulse shape parameter vs. pulse height (*right*) for a CLLB detector irradiated by the thoriated welding rods.

Each crystal is housed inside an aluminum can with a square flange. A separate aluminum can, which houses the photomultiplier tube (PMT), attaches to the crystal can. The interface between the crystal and PMT is a thin quartz window. The two cans are separable to allow for the option for another read out device to be used, if desired. The PMT used for scintillation light read out is the Hamamatsu R6231-100 - a 51-mm diameter PMT with super bialkali photocathode. The bias voltage and signal read out are done via a plug-on 14-pin voltage divider PMT base with BNC and SHV bulkhead connectors. These bases were modified from the standard 14-pin base to have 2 MΩ resistance and extra capacitance. The bias voltage to the PMT was provided by the iseg high-density, high voltage unit [6].

For data acquisition we used the Struck Innovative Systeme (SIS) model no. 3316 [7]. The SIS3316 is a 16-channel VME-based 14-bit, 250 MHz flash analog-to-digital converter (ADC). The module can operate in oscilloscope mode, which allows the user to view and store raw pulses as well as optimize the parameters of the integration (accumulator) windows, used to optimize the performance of the pulse height analysis and PSD. Using this charge integration method for PSD, we optimized the integration gates to be: short – 224 ns and long – 2 $\mu$s. The pulse shape parameter for CLLB is defined as the ratio of the short ($S$) gate to the long ($L$) gate.

Fig. 4 shows a cartoon rendering (*left*) and photograph (*right*) of the experimental set up used to perform the collimated scans. Of our 100 CLLB detectors, we selected three to scan, each one representing a group of detectors characterized by their spectral performance – high-grade (<3.5%), mid-grade (3.5% - 4.0%) and low-grade (>4.0%). The energy resolution of the specific detectors chosen were: 3.4% (high), 3.8% (mid) and 4.1% (low). All reported resolutions are measured at 662 keV (FWHM). As shown in the photograph, each detector was placed in a hut with 10-cm-thick lead on all sides (the top portion of the hut was removed for clarity in the photograph). A 5-mm-wide slit in the central portion of the hut was used as the collimated source of 2.614 MeV photons, irradiating each detector from three sides (left, right, and top). The source of 2.614 MeV gamma-ray photons was a set of thoriated welding rods. The rods are 2% thorium oxide, have a diameter of 2.4 mm, length of 15 cm, and contain 0.23 grams of thorium [8]. We completed one-hour acquisitions at one-cm increments along the entire length of the crystal, for each crystal we tested.

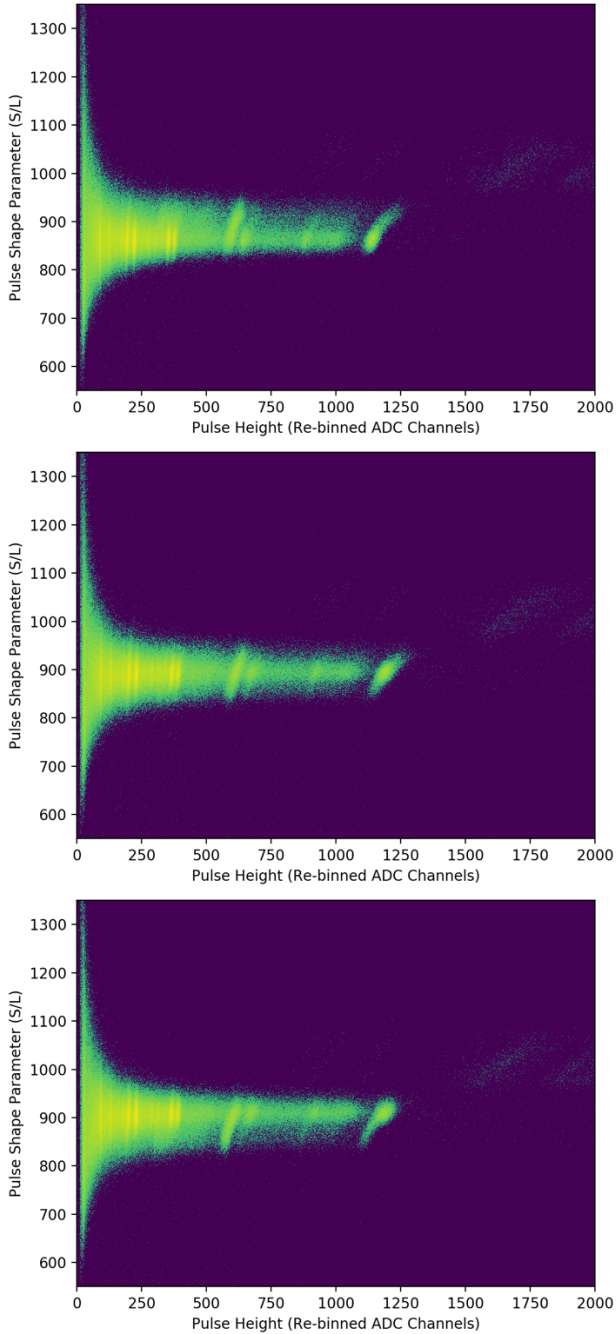

Fig. 6. 2D scatter plot of the pulse shape parameter vs. pulse height for collimated 2.614 MeV gamma rays at 2 cm (top), 5 cm (center), and 8 cm (bottom) away from the PMT-crystal interface. These data were obtained with the high-grade CLLB detector.

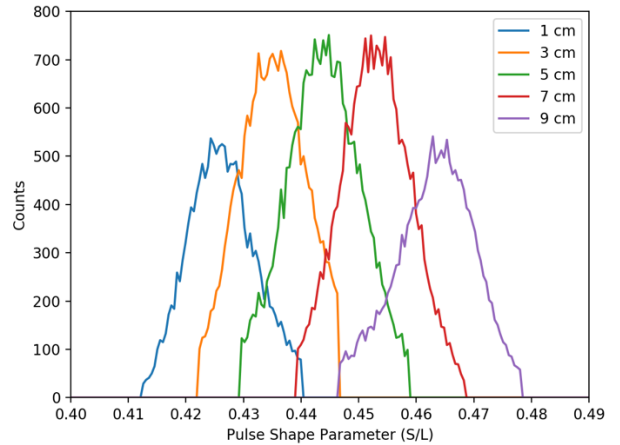

Fig. 7. 1D PSD histograms for collimated scans at 1 cm, 3 cm, 5 cm, 7 cm, and 9 cm away from the PMT-crystal interface. Data obtained with the high-grade CLLB detector.

## IV. RESULTS

Fig. 5 shows the energy spectrum (*left*) and the pulse shape parameter vs. re-binned ADC channel pulse height parameter (*right*) for the collimated thoriated rods at the center of a CLLB detector. The energy spectrum shows the 2.614 MeV full-energy peak, along with the single- and double-escape peaks at 0.511 MeV and 1.022 MeV below the full-energy peak. The other (lower) energy lines are associated with the internal gamma-ray background (from $^{138}$La) and from the thoriated rods (ranging from 238.6 keV to 968.9 keV). The 2D plot (*right*) shows the prominent 2.614 MeV gamma-ray island, well separated from the lower energy background continuum and internal alphas. Fig. 6 shows the results of the collimated scans at a few select positions along the length of the crystal, specifically 2 cm (*top*), 5 cm (*center*) and 8 cm (*bottom*) away from the PMT-crystal interface. We acquired these data using the high-grade CLLB. In these data, as well as data from the low- and mid-grade CLLB, we observed a shifting of the centroid in PSD parameter space that corresponds to the movement of the collimated source. Additionally, we observed a shift in the centroid in pulse height space, albeit to a lesser extent that the shift observed in the PSD space.

The results in Fig. 6 show multiple lines in the gamma-ray band that are either straight (vertical) or skewed in pulse shape and pulse height. The interpretation of this result is that we are observing the effects of collimation and the internal gamma-ray line. For the lines between 100 – 450 channels, these events correspond to lower-energy gamma rays that are well collimated and interact with a narrow slice of the CLLB detector. The skewed line observed around 600 channels is from the internal gamma-ray line from the $^{138}$La. Given that this line is internal and thus non-uniformly distributed along the length of the crystal, we observe the skewed line. The higher energy lines, associated with the 2.614 MeV photons, as well as the first- and second-escape peaks, appear skewed due to the collimator not being as effective at these higher energies, leading to some bleed through and thus irradiating a wider swath of the crystal than the collimated slit.

The results of the collimated scans for each CLLB are summarized in the Figs. 7 – 9. In terms of the PSD vs. collimated position, Fig. 7 shows the 1D spectra of the pulse shape parameter from five collimated points along the length of the CLLB detector, between 1 cm – 9 cm. These distributions are formed by placing 2D event selections on the pulse shape and pulse height parameters to encapsulate the 2.614 MeV gamma-ray events, then projecting the selected events onto the *y*-axis. The figure shows the magnitude of the shifting pulse shape parameter as a function of position. Next, Fig. 8 shows the centroid of the pulse shape parameter as a function of collimated scan position along the length of the crystal for the three CLLB detectors. After applying the 2D event selections described previously, we obtained the centroid location by fitting a Gaussian to each distribution. All of the detectors investigated, regardless of their intrinsic energy resolution, demonstrate the same positional-dependent characteristic.

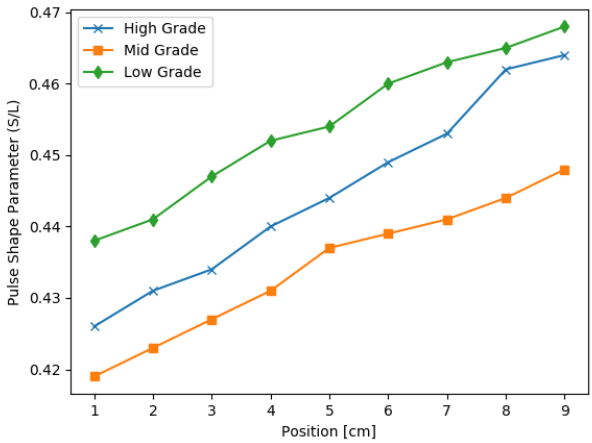

Fig. 8. PSD vs. collimated position for the high-, mid-, and low-grade CLLB detectors tested.

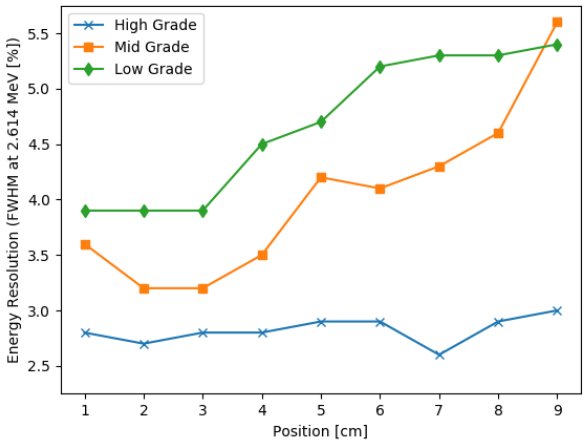

Fig. 9. Energy resolution vs. collimated position for the high-, mid-, and low-grade CLLB detectors tested.

Fig. 9 shows the results for the energy resolution (FWHM at 2.614 MeV) as a function of collimated position. The high-grade CLLB detector shows little variation in the energy resolution along the length of the crystal. Conversely, both the low- and mid-grade CLLB detectors show greater variation from one end of the crystal to the other end. The variation in the resolution along the length of the crystal contributes to the overall performance. The relative uniformity of the response of the high-grade crystal leads to the excellent performance under flood illumination. Whereas the greater variation in the resolution from one end to the other for both the low- and mid-grade crystals contributes to the overall lower performance of the crystal under flood illumination.

## V. CONCLUSIONS

For this work we found that gamma rays exhibit PSD in CLLB. This finding was unexpected and came about as a result of performing collimating scans to understand the nature of skewed thermal neutron- and alpha-particle-induced events in the crystal. Knowledge of the positional dependence of the PSD can allow for information on the interaction location within a monolithic scintillation detector. Typically, to ascertain any positional information inside a monolithic crystal, one can make the assumption that the interaction occurred at an average location along the length of said dimension, with a resolution of half the length of that dimension. For instance, consider the example of the CLLB detectors presented in this work. Along the length of the crystal (10 cm), without positional dependent PSD, the uncertainty in the interaction location is ~5 cm. With the positional-dependent PSD information, the position uncertainty can be reduced to ~1 cm.

These results have implications for monolithic detectors used in imaging systems, where poor position resolution leads to poor angular resolution. For a 2D array of detectors, one knows the ($x,y$) position of interaction based on the location of the hit detector. By knowing the position of interaction in the $z$-dimension, based on the extra PSD parameter measurement, would allow for the possibility of improved imaging performance in systems that employ these detectors.

Lastly, an understanding of the dependence of the pulse shape as a function of position could give crystal manufactures a more complete understanding of the crystal properties and the distribution of the internal contaminants and dopants.